%% file: revision.tex
\documentclass[conference]{IEEEtran}
\usepackage{times}

\usepackage{notation_commands}
\usepackage{specific_notation_commands}

\usepackage[numbers,sort&compress]{natbib}
\usepackage{multicol}
\usepackage[bookmarks=true]{hyperref}
\usepackage[nameinlink,noabbrev]{cleveref}

\usepackage{graphicx}
\usepackage{float}
\usepackage[]{units}
\usepackage{mathtools}
\usepackage{graphicx}
\usepackage[caption=false]{subfig}
\usepackage{xurl} % better URL line breaks

\def\<{ \begin{bmatrix} }
\def\>{ \end{bmatrix} }

\usepackage{xcolor}

\begin{document}

% \title{Magnetony Coilibration}
% \title{Learning and Structure for Electromagnetic Field Modeling}
% \title{Learning and Structure for Electromagnetic Field Modeling FOR REAL-TIME INVERSION}
\title{Structured Learning for Electromagnetic Field Modeling and Real-Time Inversion}

% You will get a Paper-ID when submitting a pdf file to the conference system
% \author{Author Names Omitted for Anonymous Review. Paper-ID 1127}

% \author{\authorblockN{Michael Shell}
% \authorblockA{School of Electrical and\\Computer Engineering\\
% Georgia Institute of Technology\\
% Atlanta, Georgia 30332--0250\\
% Email: mshell@ece.gatech.edu}
% \and
% \authorblockN{Homer Simpson}
% \authorblockA{Twentieth Century Fox\\
% Email: homer@thesimpsons.com}
% \and
% \authorblockN{James Kirk\\ and Montgomery Scott}
% \authorblockA{Starfleet Academy\\
% San Francisco, California 96678-2391\\
% Telephone: (800) 555--1212\\

% avoiding spaces at the end of the author lines is not a problem with
% conference papers because we don't use \thanks or \IEEEmembership

\author{\authorblockN{António Bernardes\authorrefmark{1}\authorrefmark{2},
Jasan Zughaibi\authorrefmark{1}\authorrefmark{2},
Michael Muehlebach\authorrefmark{3}, 
Bradley J. Nelson\authorrefmark{2}}
\authorblockA{\authorrefmark{1}These authors contributed equally to this work
}
\authorblockA{\authorrefmark{2}Multi-Scale Robotics Lab, ETH Zürich, 8092 Zürich, Switzerland}
\authorblockA{\authorrefmark{3}Learning and Dynamical Systems Group, Max Planck Institute for Intelligent Systems, 72076 Tübingen, Germany}
\authorblockA{Email: costaa@ethz.ch, zjasan@ethz.ch, michael.muehlebach@tuebingen.mpg.de, bnelson@ethz.ch
}
}

\maketitle

% \begingroup
% \renewcommand\thefootnote{}%
% \footnotetext{Equal contribution. Emails: \texttt{antonio@...}, \texttt{jasan@...}, \texttt{mmuehlebach@...}, \texttt{bjn@...}.}%
% \endgroup

\begin{abstract}

Precise magnetic field modeling is fundamental to the closed-loop control of electromagnetic navigation systems (eMNS) and the analytical Multipole Expansion Model (MPEM) is the current standard. However, the MPEM relies on strict physical assumptions regarding source symmetry and isolation, and requires optimization-based calibration that is highly sensitive to initialization. These constraints limit its applicability to systems with complex or irregular coil geometries. This work introduces an alternative modeling paradigm based on multi-layer perceptrons that learns nonlinear magnetic mappings while strictly preserving the linear dependence on currents. As a result, the field models enable fast, closed-form minimum-norm inversion with evaluation times of approximately \unit[1]{ms}, which is critical for high-bandwidth magnetic control. For model training and evaluation we use large-scale, high-density datasets collected from the research-grade OctoMag and clinical-grade Navion systems. Our results demonstrate that data-driven models achieve predictive fidelity equivalent to the MPEM while maintaining comparable data efficiency, and we further assess their suitability for real-time magnetic control in a closed-loop tracking experiment running at \unit[100]{Hz}. Furthermore, we demonstrate that straightforward design choices effectively eliminate spurious workspace ill-conditioning frequently reported in MPEM-based calibration. To facilitate future research, we release the complete codebase and datasets open source.

\end{abstract}

\IEEEpeerreviewmaketitle

% \section{Introduction}
% This demo file is intended to serve as a ``starter file" for the
% Robotics: Science and Systems conference papers produced under \LaTeX\
% using IEEEtran.cls version 1.7a and later.  
\input{revision/0-SupplementaryMaterial}
\input{revision/1-Introduction}
\input{revision/2-DataCollection}

\input{revision/3-ModelingMethods}
\input{revision/4-Results}

\input{revision/5-Conclusion}

\bibliographystyle{unsrtnat}
\bibliography{references}

\end{document}

%% file: revision/0-SupplementaryMaterial.tex
\section*{Supplementary material}
The complete codebase for model training, the experimental datasets, and the pre-trained models for evaluation are available at
\url{https://github.com/antonio14bernardes/CalibrationForMagneticNavigation/tree/main}.

%% file: revision/1-Introduction.tex
\label{sec:Introduction}

\section{Introduction}

\begin{figure}[htbp]
  \centering
  \subfloat[]{
    \includegraphics[width=\columnwidth]{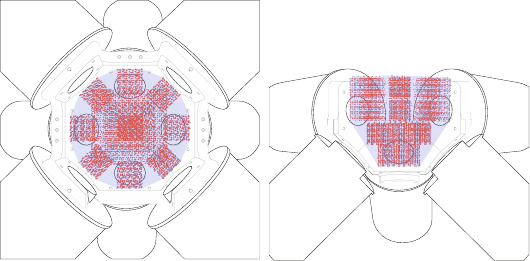}
    \label{fig:OctomagDataset}
  }
  \par
  \vspace{-4mm}
  \subfloat[]{
    \includegraphics[width=\columnwidth, trim=0cm 4cm 0cm 0cm, clip]{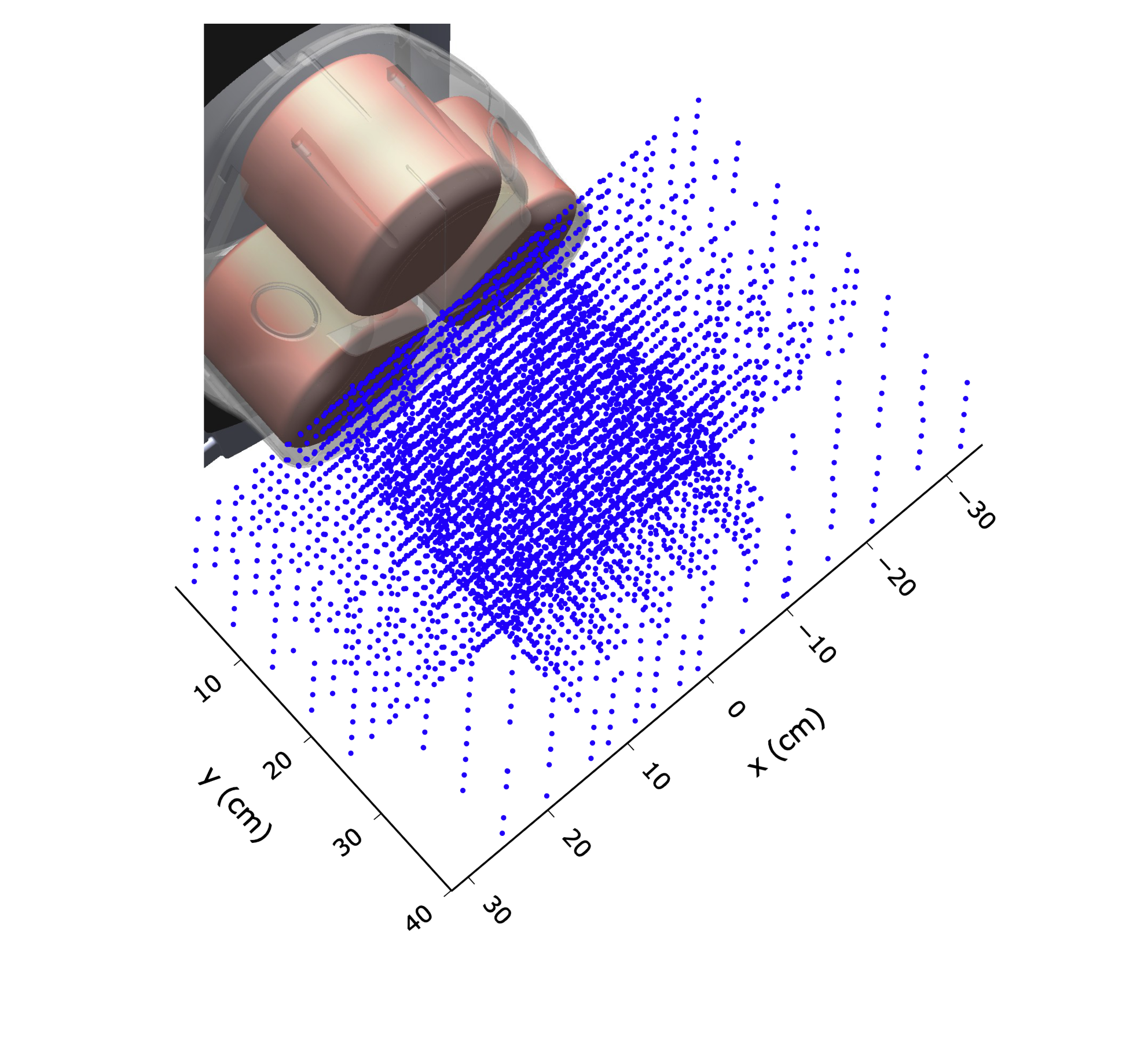}
    \label{fig:NavionDataset}
  }
      \vspace{-3mm}
  \caption{Visualization of the positions where magnetic field data has been acquired. (a) Measured positions for the Octomag system (53248 total positions). (b) Measured positions for the clinically ready Navion system (7500 total positions). To the best of our knowledge, these represent the densest datasets recorded for such systems.}
  \label{fig:CombinedMagneticDatasets}
  \vspace{-7mm}
\end{figure}

Remote magnetic manipulation is rapidly emerging as a promising technology for contactless actuation across a wide range of length scales. It enables precise, responsive control of catheters, guidewires, and endoscopes for applications ranging from telesurgery to targeted drug delivery with micro- and nanorobots \cite{fabian2025science,zhao2022tele,alex2025tele,florian2025telesurgery}. These capabilities are realized using magnetic navigation systems, which generate and shape the magnetic field using one or more magnetic sources: either permanent magnets with controllable pose \cite{pittiglio_magnetic_2019, MagneticNavigationSystemComposedOfDualPermanentMagnetsForAccuratePositionAndPostureControlOfACapsuleEndoscope} or electromagnets with controllable coil currents \cite{NavionPaper, Octomag:AnElectromagneticSystemFor5-dofWirelessMicromanipulation} (including, in some designs, moving electromagnets \cite{robomag2020_roboticeMNS}). Electromagnetic navigation systems (eMNS) arrange multiple fixed electromagnets around a workspace and, given a sufficient number and suitable placement of the coils, can synthesize a desired magnetic field at a target position by appropriately choosing the coil currents \cite{MinimumBoundsOnTheNumberOfElectromagnetsRequiredForRemoteMagneticManipulation}. Accordingly, eMNS control hinges on a field model that is not only accurate, but also computationally efficient to support closed-loop operation. This includes a forward map from coil currents to the resulting magnetic field (and, when needed, its spatial gradients), as well as an inverse model to compute currents that realize a desired output. While our focus is medical eMNS, the need for real-time field prediction and model inversion is broadly shared across electromagnetic platforms, including MRI field shaping \cite{dynMultiCoil2015}, magnetic particle imaging \cite{graser2019human}, magnetic levitation / bearing actuation \cite{miric2020dynamic}, and wireless power-transfer coil arrays \cite{sasatani2021room}. More recently, researchers in the nuclear-fusion community have discussed whether the coils in fusion facilities could support new forms of remote handling and field-aware robotics---for example, using controlled magnetic forces for partial gravity compensation in long-reach maintenance systems \cite{buckingham2016remote}.

In this work, we introduce a set of learning-based eMNS field models implemented with multilayer perceptrons (MLPs) that learn magnetic field mappings under different structure- and physics-aware representations, designed to preserve tractable, real-time inversion.  We benchmark these models against the \emph{de facto} standard, the Multipole Expansion Model (MPEM) \cite{ModelbasedCalibrationForMagneticManipulation}, and against black-box regressors (standard MLPs and gradient-boosted linear-tree models) providing an empirical comparison of model-based and model-free approaches to magnetic field modeling. We additionally demonstrate, through a closed-loop control experiment, that the proposed structured models indeed retain the real-time invertibility required for feedback control. Finally, we investigate a repeatedly noted failure mode of the MPEM---spurious ill-conditioned regions in the workspace that can destabilize inversion and control \cite{ModelbasedCalibrationForMagneticManipulation, ExpandingtheWorkspaceofElectromagneticNavigationSystemsUsingDynamicFeedbackforSingleandMultiagentControl}---and show that straightforward design choices are sufficient for mitigation. 
% In the present work, we focus on current-linear/affine forward models in the pre-saturation regime, where fields scale approximately linearly with current. This preserves superposition thus supporting tractable workspace analysis and fast closed-form minimum-norm inversion, while still accounting for highly nonlinear spatial behavior.

The analysis in this work considers two eMNS platforms: the OctoMag \cite{Octomag:AnElectromagneticSystemFor5-dofWirelessMicromanipulation} shown in \Cref{fig:CombinedMagneticDatasets}(a), a research-grade eMNS with eight coils and an approximately \unit[13]{cm}-diameter hemispherical workspace, and the Navion \cite{NavionPaper}  shown in \Cref{fig:CombinedMagneticDatasets}(b), a clinical-grade eMNS with three coils that provides magnetic control over a large, human-scale workspace. For training and evaluation, we collected, to the best of our knowledge, the largest datasets reported to date for each system, with dense sampling and broad spatial coverage across their workspaces. This coverage enables accurate modeling and assessment of spatial gradients. We release both datasets open source as supplementary material, alongside our codebase for training and a suite of pretrained models for immediate evaluation.

\subsection{Motivation for current-linear field models}

Our key emphasis is on models that are linear (or affine) in the coil currents, targeting the system’s practical pre-saturation operating regime where the electromagnet-generated field scales approximately linearly with current. Preserving current-linearity is central to eMNS practice, enabling tractable workspace and actuation-capacity analyses \cite{OnTheWorkspaceOfElectromagneticNavigationSystems, ACriticalAnalysisOfEightelectromagnetManipulationSystemsTheRoleOfElectromagnetConfigurationOnStrengthIsotropyAndAccess, ExpandingtheWorkspaceofElectromagneticNavigationSystemsUsingDynamicFeedbackforSingleandMultiagentControl}, and supporting a fast, closed-form, minimum-norm inversion via the Moore–Penrose pseudoinverse \cite{penrose_generalized_1955}. Within this regime, the contributions of individual electromagnets superpose, and an eMNS with $S$ electromagnets is typically described by a current-linear forward model: currents \(\currentsvect\in\dom{R}{S}\) produce at position \(\posP\in\dom{R}{3}\) the magnetic field vector \(\field\in\dom{R}{3}\) as
$\field=\func{\actuation_{\field}}{\posP}\currentsvect,
$
where $\func{\actuation_{\field}}{\posP}\in\dom{R}{3\times S}$ is commonly referred to as the field actuation matrix.

In the present study, we augment the standard linear formulation with a position-dependent affine term, leading to
\begin{equation}
    \field=\func{\actuation_{\field}}{\posP}\currentsvect+\func{\field_0}{\posP},
    \label{eq:AffineMap}
\end{equation}
which preserves the convenience of workspace analysis and inversion, while capturing background contributions such as sensor-current biases (e.g., from miscalibration) and remanent magnetization within the workspace. Note that although \eqref{eq:AffineMap} is affine in the currents, both $\func{\actuation_{\field}}{\posP}$ and $\func{\field_0}{\posP}$ may vary strongly/nonlinearly with position. In our learning-based approaches, we learn these terms either directly, or indirectly via a physics-informed scalar potential parameterization.% and benchmark them to an affine representation of the MPEM. 

% The MPEM is itself formulated under the current-linearity assumption,  using a model-based multipole expansion to capture the field’s nonlinear spatial dependence, and we augment it with a position-dependent affine offset as in \eqref{eq:AffineMap}. Similarly, we introduce learning-based models that leverage the current-affinity by learning position-dependent affine mappings (i.e. the parameters of \eqref{eq:AffineMap}), either directly or through a physics-informed scalar potential-based representation, rather than only regressing the field as a generic function of \((\posP,\currentsvect)\).

\subsection{Why not rely on nonlinear forward models?}
Prior work has applied learning-based models to eMNS primarily to extend accuracy beyond the linear regime, where higher currents drive ferromagnetic cores toward saturation and the MPEM's linearity assumption, along with its modeling performance, breaks down. In particular, random forests, MLPs, and convolutional neural networks have been shown to achieve strong overall predictive accuracy across operating conditions, including regimes where saturation effects are non-negligible \cite{ModelingElectromagneticNavigationSystems, ModelingElectromagneticNavigationSystemsForMedicalApplicationsUsingRandomForestsAndArtificialNeuralNetworks, ernst_physics-informed_2025}. However, these gains can come with a trade-off: models optimized for global nonlinear fidelity do not necessarily preserve high accuracy in the near-linear operating range that is most relevant for fast, well-conditioned inversion. Nevertheless, nonlinear models have seen limited adoption in remote magnetic motion control, in part because they forfeit the algebraic structure available in the linear regime: the inverse problem is no longer a pseudoinverse and instead requires more expensive inversion techniques such as iterative nonlinear optimization \cite{ModelingElectromagneticNavigationSystems, MagneticMethodsInRobotics}, formulation as a differential control problem \cite{MagneticMethodsInRobotics}, or a separately learned inverse map \cite{ernst_physics-informed_2025}.

In addition, many eMNS applications operate in highly dynamic settings that demand high-bandwidth control \cite{zughaibi_dynamic_2025, ExpandingtheWorkspaceofElectromagneticNavigationSystemsUsingDynamicFeedbackforSingleandMultiagentControl, singh_remote_2025, berkelman_magnetic_2013, miyasaka_magnetic_2014, berkelman_multiple_2023}, making inverse-model latency a primary concern \cite{xu_fpga-based_2018}. Iterative inversion schemes for nonlinear forward models are therefore difficult to accommodate at the required update rates.

\subsection{Why not learn the inverse map directly?}
While the inverse-MLP in \cite{ernst_physics-informed_2025} can, in principle, provide low-latency inference, fundamentally, learning an inverse map from fields (and gradients) to currents cannot, in general, guarantee energy-optimal actuation, even when regularized toward minimum-norm currents given a field/gradient target. Since many feedback control algorithms specify objectives in motion space (torques/forces), optimizing an inverse in this intermediate space (fields-to-currents) independently of the full composition does not yield a globally optimal solution to the end-to-end inverse relevant for control. The resulting currents can differ by an order of magnitude as shown in \cite{ExpandingtheWorkspaceofElectromagneticNavigationSystemsUsingDynamicFeedbackforSingleandMultiagentControl}. Furthermore, practical current inversion may need to be modified online to account for saturation, power-supply limits, and temperature-dependent weighting \cite{pourkandMagneticActuationStationary2020}, which a fixed learned inverse cannot naturally accommodate.

In contrast, because the mapping from field/gradients to magnetic torque/force is linear, an affine field model preserves affinity from currents to the control input, enabling fast, closed-form, minimum-norm inversion of the end-to-end mapping using a computationally inexpensive pseudoinverse operation. 

\subsection{Outline}
The remainder of the paper is structured as follows. \Cref{sec:DataCollection} introduces the experimental platforms and details the data collection procedure, including an empirical assessment of linearity and outlier removal. \Cref{sec:ModelingMethods} describes the proposed learning-based architectures alongside the benchmark models. \Cref{sec:Results} presents a comprehensive empirical evaluation of the models, covering predictive accuracy, data efficiency, physical consistency, computational performance, and performance in a closed-loop control task, while also investigating the sources of residual error and examining MPEM-induced ill-conditioning in field actuation maps. Finally, \Cref{sec:Conclusion} summarizes the findings and discusses implications for real-time control.

%% file: revision/2-DataCollection.tex
\section{Data collection}
\label{sec:DataCollection}

Reliable learning-based field models require consistent measurements across currents and space; we therefore detail our acquisition protocol and the verification/cleaning steps applied to both datasets.

\subsection{Experimental platforms}

\subsubsection{OctoMag}
To collect magnetic-field data on the OctoMag, we used a $4\times4\times4$ array of Melexis\textsuperscript{\textregistered} MLX90395 triaxial high-field magnetometers with \unit[5]{mm} sensor spacing. The array was mounted in a rigid fixture and positioned at a wide range of poses, yielding a dense dataset spanning the full workspace with an effective spatial sampling of approximately \unit[1.7]{mm}, as shown in \Cref{fig:CombinedMagneticDatasets}(a). To ensure accurate pose measurements, the cube was tracked with an infrared motion-capture system using reflective markers.

We collect data with currents in the range $\pm \mathrm{\unit[4]{A}}$. Since we focus on the linear regime where superposition holds, each coil’s contribution can be modeled independently. Accordingly, to streamline data collection, we ramped one coil at a time from -4 to \unit[4]{A} in \unit[1]{A} increments while holding all other coils at zero current.

\subsubsection{Navion}
For the Navion, we used a separate, larger sensor array arranged in a $5\times5\times5$ grid of Allegro\textsuperscript{\textregistered} A1301 magnetometers with \unit[5]{cm} spacing. The array was manually positioned at a large set of positions throughout the workspace, with denser sampling in regions closer to the magnetic sources. As in the OctoMag procedure, the sensor array was tracked using an infrared motion-capture system. The spatial extent of this dataset is seen in \Cref{fig:CombinedMagneticDatasets}(b).

Data were acquired using coil currents from 0 to \unit[30]{A} in \unit[5]{A} steps, with the same per-coil sweep strategy as for the OctoMag.

\subsection{Linearity in currents}

\begin{figure}[h]
  \centering \includegraphics[width=\columnwidth]{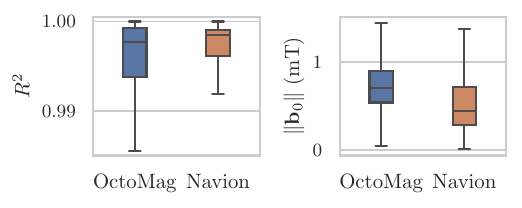}
  \vspace{-8mm}
  \caption{Evaluation of magnetic field linearity through boxplots of fit quality ($R^2$) and field offset magnitudes ($\|\field_0\|$) for the OctoMag and Navion. The consistently high $R^2$ values confirm that both systems operate within the linear-current regime over the tested ranges, while the nonzero offsets indicate persistent constant field contributions.}
  \label{fig:Linearity}
  \vspace{-4mm}
\end{figure}

For both systems, we verified that the collected current ranges fall within their corresponding linear regime. For each sampled position, we fit an affine model relating the measured magnetic field to
the commanded current in each electromagnet. The distributions of $R^2$ values from the per-position affine fits and those of the affine intercept magnitudes are shown in \Cref{fig:Linearity}. The consistently high $R^2$ values support linear behavior over the tested ranges, while the intercept magnitudes indicate a persistent constant field offset, likely originating from a miscalibration of the current sensors.

% \begin{figure*}[h]
%   \centering \includegraphics[width=\textwidth]{figures/linearity_check.pdf}
%   \vspace{-8mm}
%   \caption{Validation of magnetic field linearity. (a) Boxplots of fit quality ($R^2$) and field offset magnitudes ($\|\field_0\|$) for the OctoMag and Navion. (b--c): Representative magnetic field measurements versus commanded current for the OctoMag (b) and Navion (c), demonstrating the affine relationship between current and field generation in both systems.}
%   \label{fig:Linearity}
%   \vspace{-5mm}
% \end{figure*}

% For each system, we verified that the current ranges used for data collection fall within their corresponding linear regime. For the OctoMag, prior to the full workspace sweep we placed the sensor array near the workspace center and ramped all coils simultaneously from minimum to maximum current in \unit[0.2]{A} steps. For neighboring coils we inverted the commanded polarity so that adjacent cores were magnetized constructively, promoting saturation and therefore a conservative linearity check. For each of the 64 measured locations, we fit an affine model (with intercept) relating the measured magnetic field to the commanded current. For the Navion, the same analysis was performed \emph{post hoc} on a subset with 100 positions randomly sampled from the full dataset, and therefore with only coil active at a time.

% The distributions of $R^2$ values from the per-position affine fits are shown in \Cref{fig:Linearity}(a), and those of the affine intercept magnitudes are shown in \Cref{fig:Linearity}(b). The consistently high $R^2$ values support linear behavior over the tested ranges, while the intercept magnitudes indicate a persistent constant field offset, likely originating from a miscalibration of the current sensors.

% \begin{figure*}[h]
%   \centering \includegraphics[width=\textwidth]{figures/linearity_check.pdf}
%   \caption{Not final plot. Just to see how it will look}
%   \label{fig:Linearity}
% \end{figure*}

\subsection{Outlier removal}
To remove anomalous measurements, we use RANSAC \cite{fischler_random_nodate}, a robust estimator for structured models commonly used for outlier detection in the field of computer vision. Specifically, we exploit the observed affine-in-currents substructure and
 the per-coil sweep collection strategy of our data: for each position and active coil, we fit an affine model using RANSAC and classify as outliers those samples whose residuals exceed the inlier tolerance of the consensus fit.
 
A key design choice is the inlier criterion. A criterion based on the absolute residual, $\norm{\field-\pred{\field}}$, tends to reject large-magnitude samples that are accurate in relative terms, whereas a purely relative residual, $\norm{\field-\pred{\field}}/\norm{\field}$, over-penalizes smaller measurements---this is undesirable because low-magnitude samples encode the field's rapid decay as the distance from the source increases, an important concern for modeling. We therefore use a scale-aware relative residual,
\begin{equation}
r \;=\; \frac{\norm{\field-\pred{\field}}}{\max(\norm{\field},\,\fieldscalar_T)},
\end{equation}
with a denominator floor of $\fieldscalar_T$  that prevents the relative error from being over-amplified in the low-field regime. We set the floor magnitude to the first quartile of the $\norm{\field}$ distribution, namely, $\fieldscalar_T=\unit[2.66]{mT}$ for the OctoMag dataset and $\fieldscalar_T=\unit[3.27]{mT}$ for the Navion dataset. A measurement is classified as an outlier if $r > 20\%$.

This procedure flagged approximately 5.03\% of the OctoMag dataset as outliers and approximately 3.11\% for the Navion. Note that, since the criterion is based on consistency with currents, it will not detect samples whose field direction is systematically corrupted (e.g., due to a pose-tracking error) but that remain affine in the current command.

%% file: revision/3-ModelingMethods.tex
\section{Modeling methods}
\label{sec:ModelingMethods}

Magnetic fields used for magnetic manipulation are well described by the quasi-static Maxwell equations. In the workspace, assumed to be current-free, these reduce to:
\begin{equation}
\nabla \cdot \field = 0,
\qquad
\nabla \times \field = \vect{0}.
\label{eq:Maxwell}
\end{equation}
Thus, the workspace field is both divergence-free and curl-free. In eMNS control, a scalar potential $\Phi$ is commonly used to express the field as
\begin{equation}
\field = -\nabla \Phi.
\label{eq:BFromPotential}
\end{equation}
Substituting \eqref{eq:BFromPotential} into the divergence-free condition yields Laplace's equation:
\begin{equation}
\nabla^2 \Phi = 0.
\label{eq:LaplacePhi}
\end{equation}

Conditions \eqref{eq:Maxwell}--\eqref{eq:LaplacePhi} provide a useful set of physical constraints for model formulation and physical-consistency evaluation of model-free methods.

\subsection{The Multipole Expansion Model}
\label{subsec:MPEM}
The MPEM, introduced in \cite{ModelbasedCalibrationForMagneticManipulation}, models each magnetic source by an azimuthally symmetric scalar potential $\Phi$. Under this formulation, and considering a workspace external to the electromagnets, Laplace's equation \eqref{eq:LaplacePhi} admits the truncated multipole expansion

\begin{equation}
\Phi(r,\cos\theta)=\sum_{n=1}^{N}\frac{B_n}{r^{n+1}}P_n(\cos\theta),
\label{eq:MPEM_Expansion}
\end{equation}
where $N$ is the order of approximation, $P_n(\cdot)$ denotes the $n$-th Legendre polynomial, $B_n$ is the corresponding multipole coefficient, $r$ is the distance to the evaluation point, and $\theta$ is the polar angle from the source’s zenith direction (i.e., its symmetry axis). The MPEM therefore satisfies Maxwell's equations \eqref{eq:Maxwell} by construction.

As mentioned above, the MPEM is formulated under the assumption of linearity in currents, hence the field contribution of a single source is written as $\func{\field}{\posP} = \current\,\func{\tilde{\field}}{\posP}$, where the unit current field contribution $\tilde{\field}$ follows from \eqref{eq:BFromPotential}; for brevity, we refer the reader to \cite{ModelbasedCalibrationForMagneticManipulation} for the resulting closed-form expression. To account for cross-magnetization, the MPEM augments each of the $S$ electromagnets with $F$ sources that model induced magnetization in nearby ferromagnetic components---namely the cores of the surrounding electromagnets.
Additionally, to accommodate the non-negligible constant field offset observed in \Cref{sec:DataCollection}, we augment the model by adding one source per electromagnet, each subject to a constant unit-current. The resulting field produced by currents vector $\currentsvect = \left[\,\current_1,\ldots, \current_S\,\right]\transposed$ at position $\posP$ is:
\begin{equation}
\func{\field}{\posP, \currentsvect}
=
\sum_{s=1}^{S}
\current_s
\left(\func{\tilde{\field}_s}{\posP} +
\sum_{f=1}^{F}\func{\tilde{\field}_{s,f}}{\posP}
\right)
+
\func{\field_{s,0}}{\posP}
,
\label{eq:MPEM_total}
\end{equation}
where $\tilde{\field}_{s}$ and $\tilde{\field}_{s,f}$ are the unit-current field contributions of the $s$-th electromagnet by direct current excitation and through induced magnetization of ferromagnetic body $f$, respectively, and $\field_{s,0}$ (in $\mathrm{T}$) is the offset term. Collecting the per-electromagnet contributions in \eqref{eq:MPEM_total} yields the affine map \eqref{eq:AffineMap}.

Calibration then amounts to estimating the $B_n$ coefficients and the poses (location and zenith direction) of each source, iteratively solving a non-linear least squares problem using the Levenberg--Marquardt algorithm \cite{ModelbasedCalibrationForMagneticManipulation}.

\subsection{The ActuationNet}
We have repeatedly emphasized the utility of the linear (or affine) structure applicable in the linear regime. Accordingly, we introduce the ActuationNet, an MLP $f_\mathrm{AN}$ that takes as input a position $\posP\in\dom{R}{3}$ in the workspace and predicts the local field actuation matrix and affine offset of \eqref{eq:AffineMap}:
\begin{equation}
(\pred{\actuation_{\field}}, \pred{\field}_0) = f_{\mathrm{AN}}(\posP),
\qquad
\func{\pred{\actuation_{\field}}}{\posP}\in\dom{R}{3\times S},\ 
\func{\pred{\field}_0}{\posP}\in\dom{R}{3}.
\end{equation}
The ActuationNet and all MLPs introduced below use $\tanh$ as the activation function in all hidden layers.

\subsection{The PotentialNet}
While the MPEM, by construction, preserves the curl- and divergence-free conditions, the ActuationNet is free to learn an arbitrary affine mapping and thus does not necessarily respect Maxwell’s equations \eqref{eq:Maxwell}. As an alternative, we introduce the PotentialNet, an MLP $f_{\mathrm{PN}}$ that maps a workspace position $\posP\in\dom{R}{3}$ to a set of scalar potentials: one unit-current potential per coil and a bias potential:
\begin{equation}
(\pred{\boldsymbol{\phi}},\pred{\phi}_0)
=
\func{f_{\mathrm{PN}}}{\posP},
\qquad
\func{\pred{\boldsymbol{\phi}}}{\posP}\in\dom{R}{S},\ 
\func{\pred{\phi}_0}{\posP}\in\dom{R}{},
\label{eq:PotentialNet_out}
\end{equation}
which define the predicted scalar potential
\begin{equation}
\func{\pred{\Phi}}{\posP,\currentsvect}
=
\func{\pred{\boldsymbol{\phi}}}{\posP}^{\transposed}\currentsvect+\func{\pred{\phi}_0}{\posP}.
\label{eq:PotentialNet_Phi}
\end{equation}
Taking the gradient of \eqref{eq:PotentialNet_Phi} with respect to $\posP$ yields a predicted affine map of the form \eqref{eq:AffineMap} which is curl-free by construction:
\begin{equation}
\begin{aligned}
\func{\pred{\actuation_{\field}}}{\posP}
&= -\func{\nabla\pred{\boldsymbol{\phi}}}{\posP}
\in\dom{R}{3\times S},\\
\func{\pred{\field}_0}{\posP}
&= -\func{\nabla\pred{\phi}_0}{\posP}
\in\dom{R}{3}.
\end{aligned}
\label{eq:PotentialNet_Affine}
\end{equation}
Note that this formulation guarantees a curl-free predicted field, but it does not enforce divergence-freeness. While, in principle, a learning-based method could enforce both curl- and divergence-free conditions by learning a parameterization over harmonic-gradient basis functions, such a choice would effectively amount to learning the parameters of a reduced, MPEM-like model class (a harmonic potential with a prescribed basis), and would therefore blur the distinction between model-based and model-free approaches that we seek to benchmark. Alternatively, we could encourage divergence-freeness during training via a Laplacian-based regularization term \cite{ernst_physics-informed_2025}, but we do not apply such regularization in the present work.

\subsection{Benchmark mappings}
\label{sec:BenchmarkMappings}

As additional baselines, we evaluate two models that directly learn the current-to-field map $\pred{\field}=\func{f}{\posP,\currentsvect}$, to quantify the value of the reduced representations used by the ActuationNet and the PotentialNet. In contrast to our structured models, these direct baselines impose no explicit assumption of  affinity in currents. We consider the DirectNet, an MLP, and the DirectGBT, a gradient-boosted linear-tree regressor (i.e., each leaf predicts a $\pred{\field}$ contribution via a local affine function of the inputs). Because data are collected via per-coil sweeps, the dataset contains only one-hot current vectors, so these direct models may extrapolate poorly to simultaneous multi-coil actuation; we therefore consider them strictly as benchmarks in the current setting, not as deployable models without additional multi-coil training.

% \subsection{Learning-based model structure}
% For each neural model, we consider three network sizes with hidden-layer widths $(256,256)$, $(256,256,256)$, and $(512,512,512)$, using $\tanh$ activations in all hidden layers and a linear output layer. We also train three DirectGBT instances with a maximum of 32, 64, and 128 leaves per tree. Finally, for the MPEM, we consider first (dipole), second (quadrupole), and third (octopole) order approximations.

\subsection{Training}
\label{sec:Training}
All models are trained to minimize the mean-squared error (MSE) in the magnetic field predictions,
\begin{equation}
\mathcal{L}_{\mathrm{MSE}}=\frac{1}{N}\sum_{i=1}^{N}\norm{\field_i-\pred{\field}_i}^{2},
\end{equation}
requiring the output of the ActuationNet to be converted with \eqref{eq:AffineMap} and that of the PotentialNet with \eqref{eq:PotentialNet_Affine} and \eqref{eq:AffineMap} prior to loss evaluation.

The full datasets are randomly split into 70\% training, 15\% validation, and 15\% test sets based on the positional features. This means that for any fixed position $\posP$, all corresponding samples across all applied currents $\currentsvect$
remain entirely within one split (train, validation, or test). Although a conventional random split over the full $(\posP,\currentsvect)$ feature set may appear appealing, it is inappropriate for the affine models (MPEM, ActuationNet, PotentialNet), whose outputs depend only on position. A sample-wise split would distribute identical positions across different splits, leading to information leakage and, moreover, may leave the training set with an insufficient diversity of current values at a given position to reliably estimate the affine parameters (slope and intercept).

We train the MPEM as described in Section \ref{subsec:MPEM} and stop when the validation RMSE, i.e., $\sqrt{\mathcal{L}_{\mathrm{MSE}}}$, changes by less than $\unit[10^{-4}]{mT}$ between successive iterations. Neural networks are trained with the Adam optimizer \cite{kingmaAdamMethodStochastic2017} (learning rate $10^{-3}$) and stopped when the validation RMSE fails to improve by at least $\unit[10^{-3}]{mT}$ over 10 consecutive epochs. DirectGBT training is stopped after 200 boosting iterations without validation improvement, using a shrinkage factor of 0.1 for each new tree. For all models, we retain the parameter set that achieves the lowest validation RMSE.

\subsection{Gradient modeling}
Spatial derivatives of the predicted field provide a useful diagnostic for physical consistency with Maxwell’s equations \eqref{eq:Maxwell}. More importantly, obtaining spatial derivatives of  the magnetic fields in real-time is critical for magnetic control, since magnetic forces depend on field gradients. In eMNS control, the constraints in \eqref{eq:Maxwell} allow for a reduced representation of the full $3\times 3$ spatial Jacobian. By convention, we define the gradient vector as
\[
\fieldgrad
\coloneq \brvect{
\frac{\partial \fieldscalar_x}{\partial x},
\frac{\partial \fieldscalar_x}{\partial y},
\frac{\partial \fieldscalar_x}{\partial z},
\frac{\partial \fieldscalar_y}{\partial y},
\frac{\partial \fieldscalar_y}{\partial z}
}\transposed.
\]
Affinity in coil currents is preserved under positional differentiation. Stacking the field and reduced gradient yields
\begin{equation}
\begin{bmatrix}
\field\\
\fieldgrad
\end{bmatrix}
=
\begin{bmatrix}
    \actuation_{\field}(\posP)\\
    \actuation_{\fieldgrad}(\posP)
\end{bmatrix}
\,\currentsvect
+
\begin{bmatrix}
\func{\field_0}{\posP}\\
\func{\fieldgrad_0}{\posP}
\end{bmatrix},
\label{eq:FullMap}
\end{equation}
where $\func{\actuation_{\fieldgrad}}{\posP}\in\dom{R}{5\times S}$ is referred to as the gradient actuation matrix.

Although our datasets do not include direct measurements of magnetic field gradients, we can still obtain gradient estimates by exploiting the structure of each model. For the MPEM, spatial derivatives follow analytically from its closed-form expression. For the neural models, using $\tanh$ as the activation function, the forward mappings are inherently continuous and smooth, with gradients being obtained via automatic differentiation. For the DirectGBT, gradients are obtained by summing the derivatives of the active linear leaves across trees, yielding a piecewise-constant Jacobian, undefined only on split boundaries.

%% file: revision/4-Results.tex
\section{Results}
\label{sec:Results}
In this section, we evaluate the proposed structured learning-based models against the physics-based MPEM and unstructured learning-based benchmarks (DirectNet, DirectGBT) using the collected datasets. We evaluate predictive accuracy relative to model complexity and data density, analyze the sources of residual error via geometric sensitivity analysis, assess physical consistency with Maxwell’s equations, examine computational efficiency for real-time control, and demonstrate the applicability of the proposed models in a closed-loop tracking task. Finally, we propose practical methods to mitigate the MPEM-induced workspace ill-conditioning.

\begin{figure*}
    
  \centering \includegraphics[width=\textwidth]{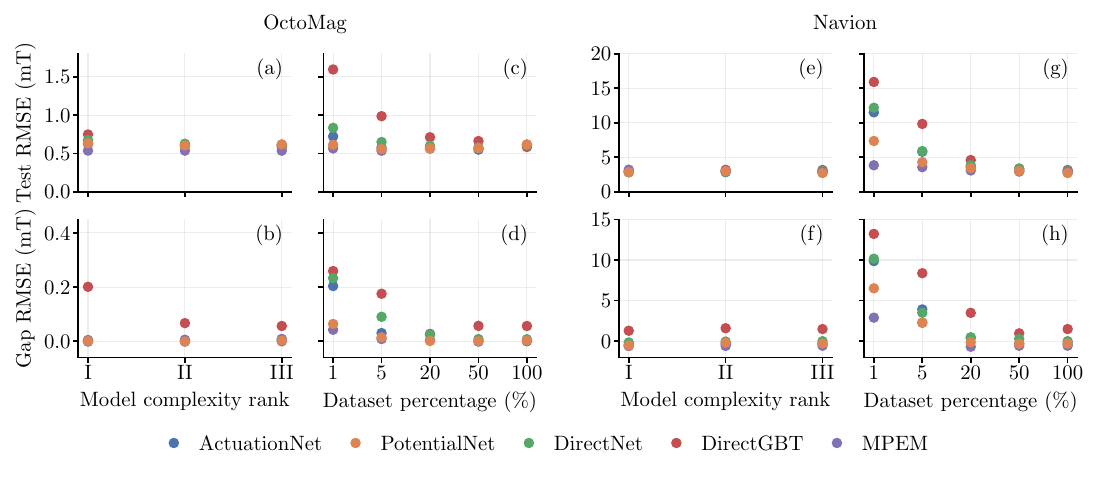}
  \vspace{-10mm}
  \caption{Field prediction accuracy, model complexity, and data-efficiency analysis for the OctoMag and the Navion. Panels (a) and (e) show test-set RMSE as a function of model complexity rank, while (b) and (f) report the corresponding generalization gap, defined as $\mathrm{RMSE}_{\mathrm{test}}-\mathrm{RMSE}_{\mathrm{train}}$. Test RMSE is largely insensitive to model complexity, indicating an error floor not primarily driven by model capacity. Panels (c) and (g) show test-set RMSE as a function of dataset percentage for the highest-capacity variant of each model class, and (d) and (h) show the corresponding generalization gap. For the OctoMag, the structured models, especially the PotentialNet, remain comparatively robust under reduced data availability, whereas unstructured baselines show stronger overfitting. For the Navion, where the baseline positional density is lower, model degradation occurs earlier in the sweep. The PotentialNet remains the most robust learning-based model, although it does not reach MPEM-level robustness at the lowest density.}
  \label{fig:FieldPred}
  \vspace{-5mm}
\end{figure*}

% \begin{figure*}
%   \centering \includegraphics[width=\textwidth]{figures/rmse_suite.pdf}
%   \vspace{-8mm}
%   \caption{Field prediction performance and data efficiency analysis. (a) Test RMSE versus model complexity rank (I, II, III). All model classes converge to an error floor of approximately 0.5 mT, demonstrating that performance is largely insensitive to increased model capacity. (b) Generalization gap (difference between test RMSE and train RMSE) versus complexity rank. The gap remains near zero for most methods, indicating minimal overfitting with the full dataset. (c) Test RMSE as a function of dataset percentage. Predictive accuracy remains stable for most models down to 20\% of the training data. (d) Generalization gap as a function of dataset percentage. At low data regimes (1–5\%), unstructured baselines (DirectNet, DirectGBT) exhibit significant overfitting, while structured models maintain a smaller generalization gap.}
%   \label{fig:FieldPred}
%   \vspace{-5mm}
% \end{figure*}

\subsection{Field prediction and model complexity}
\label{subsec:FieldPrediction}

We first evaluate field prediction accuracy across model classes. As shown in \Cref{fig:FieldPred}(a) and (e), all methods achieve comparable test-set performance, converging to an RMSE of nearly \(0.5\)~mT for the OctoMag dataset and \(3\)~mT for the Navion dataset. To analyze the effect of model size/complexity, we define three complexity ranks (I, II, III) within each model class. For neural models, the ranks correspond to hidden-layer structures with increasing neuron counts, i.e. \((256,256)\), \((256,256,256)\), and \((512,512,512)\), respectively. For gradient-boosted tree models, ranks correspond to increasing per-tree leaf limits (32, 64, and 128 leaves). For the MPEM baseline, ranks correspond to increasing approximation/multipole order (first/dipole, second/quadrupole, third/octopole). Using these definitions, \Cref{fig:FieldPred}(a) and (e) confirm that test RMSE is largely insensitive to complexity rank across model classes, suggesting an error floor that is not driven by model capacity.

Finally, to quantify overfitting, we report the generalization gap (Gap RMSE), defined as \(\mathrm{RMSE}_{\text{test}}-\mathrm{RMSE}_{\text{train}}\), in \Cref{fig:FieldPred}(b) and (f). With the exception of DirectGBT, which shows a higher tendency to overfit, the gap remains near zero for all models, indicating minimal overfitting; moreover, this behavior is again insensitive to model capacity.

\subsection{Dependence on positional resolution}
\label{subsec:DataAmount}
To evaluate how performance degrades with reduced data resolution, we retrain the highest-capacity variant (complexity rank III) of each model class using progressively smaller random subsamples of the training and validation sets, while keeping the test set fixed. Given that high-resolution acquisition is resource-intensive, this analysis aims to identify the minimum sampling density required to maintain predictive fidelity. Subsampling is performed at the positional level as detailed in \Cref{sec:Training}. We report, for each system, the test-set RMSE in \Cref{fig:FieldPred}(c) and (g) as well as the generalization gap in \Cref{fig:FieldPred}(d) and (h) as a function of dataset percentage.

For the OctoMag (\Cref{fig:FieldPred}(c) and (d)), test RMSE and generalization remain largely stable under data reduction down to \(20\%\). At \(5\%\), both unstructured baselines show clearer signs of overfitting (increased generalization gap), whereas the structured models retain a smaller gap. At \(1\%\), the ActuationNet begins to noticeably overfit while the PotentialNet retains performance approaching MPEM-level robustness, suggesting that imposing model structure is beneficial for mitigating overfitting without sacrificing test RMSE.

For the Navion (\Cref{fig:FieldPred}(g) and (h)), similar trends are observed, although, given the lower baseline positional density of the Navion dataset, performance degradation occurs earlier in the data reduction sweep. As with the OctoMag results, PotentialNet remains the most robust learning-based model under reduced data availability. At the lowest data percentage, however, it does not reach the MPEM's performance.

% Overall, imposing model structure appears beneficial for mitigating overfitting without sacrificing test RMSE: the PotentialNet degrades less than the ActuationNet, and the MPEM is the least sensitive model to data reduction. Notably, however, the structured learning-based models remain highly competitive across the entire data-efficiency sweep, approaching MPEM-level robustness at low sampling densities.

\subsection{RMSE floor and uncertainty analysis} 
As shown in \Cref{fig:FieldPred}, all model classes, even with increasing capacities, converge to a similar RMSE of approximately \unit[0.5]{mT} for the OctoMag dataset and \unit[3]{mT} for the Navion when provided sufficient data. This convergence suggests  error floors determined not by model capacity, but by physical limitations. Accordingly, we now investigate the hypothesis that this residual error is dominated by geometric uncertainties---specifically, small deviations in sensor position and orientation.

The sensor array used for the OctoMag comprises four flexible PCBs mounted in a 3D-printed fixture that also carries the motion-capture markers, while for the Navion each sensor is manually assembled onto the fixture. Manufacturing and assembly tolerances inevitably introduce mechanical deviations, causing fixed offsets between the actual sensor poses and the marker-defined frame. Although these geometric offsets are static, they induce varying field errors as the arrays are reoriented throughout the workspace during data collection. We model this geometric uncertainty as an isotropic translational error $\delta\posP\sim \mathcal{N}(\vect{0},\sigma_p^2\identity)$ and a small-angle rotational error $\delta\theta\sim \mathcal{N}(\vect{0},\sigma_\theta^2\identity)$, where $\identity\in\dom{R}{3\times3}$ is the identity matrix. Using a first-order Taylor expansion, the resulting field-error covariance $ \Sigma \in \dom{R}{3\times3}$ for a given sample is:
\begin{equation}
    \Sigma=\sigma_p^2\,\nabla\field\,\nabla\field\transposed+ \sigma_{\theta}^2\,\skewm{\field}\,\skewm{\field}\transposed,
\end{equation}
where $\skewm{\field}$ is the skew-symmetric cross-product matrix of the field vector. Using a dipole-order MPEM to approximate the gradients, we compute the expected RMSE induced by these tolerances over the $N$ samples of each test set:
\begin{equation}
    \mathrm{RMSE}_{\mathrm{floor}}=\sqrt{\frac{1}{N}\sum_{i=1}^{N} \mathrm{trace}\!\left(\Sigma_{i}\right)}.
\end{equation}

\begin{figure}
  \centering
  \includegraphics{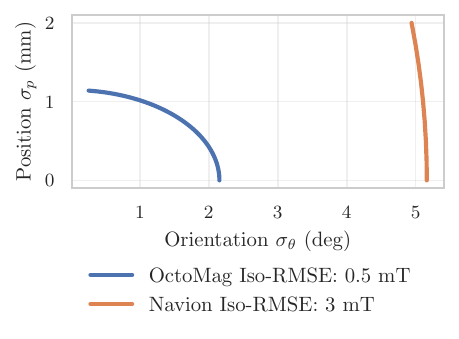}
  \vspace{-6mm}
  \caption{Position–orientation uncertainty combinations that reproduce the observed RMSE floor under the first-order geometric uncertainty model. Each curve shows the $\mathrm{RMSE}_{\mathrm{floor}}$ iso-contour corresponding to the empirical error floor of the respective system: \unit[0.5]{mT} for the OctoMag and \unit[3]{mT} for the Navion. This analysis demonstrates that sub-millimeter position error and degree-scale orientation error reproduce the observed error floors. The larger uncertainty levels required for the Navion are consistent with its manually assembled sensor array, whereas the OctoMag array is PCB-based and therefore more geometrically certain.}
  \label{fig:Sensitivity}
  \vspace{-6mm}
\end{figure}

The \unit[0.5]{mT} and \unit[3]{mT} iso-contours for the OctoMag and Navion, respectively, in \Cref{fig:Sensitivity} demonstrate that realistic combinations of sub-millimeter position error and degree-scale orientation error reproduce the observed error floors. This supports the hypothesis that geometric uncertainty, rather than model capacity, is the dominant source of residual error.

\subsection{Physical consistency}

Beyond prediction accuracy, we evaluate each model’s physical consistency by checking adherence to the quasi-static Maxwell's equations \eqref{eq:Maxwell}. Focusing on the highest-capacity variants (rank III) trained on the full datasets, we compute the divergence and curl magnitude using the spatial derivatives on each test set. The MPEM is excluded from this analysis as it satisfies both conditions in \eqref{eq:Maxwell} by construction.

\begin{figure}[h]
  \centering
  \vspace{-4mm}
  \includegraphics{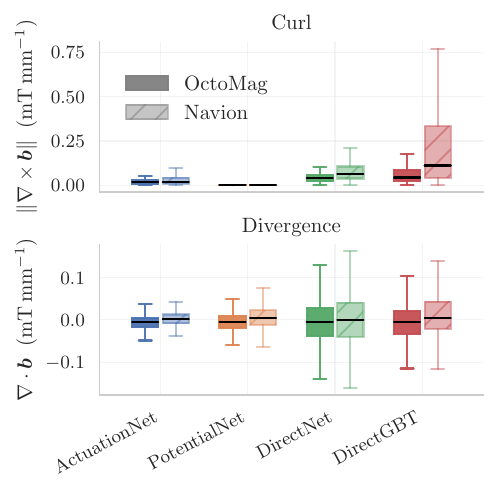}
  \vspace{-5mm}
  \caption{Evaluation of physical consistency through boxplots showing the distribution of curl magnitude (top) and divergence (bottom) computed on each system's test set. The structured learning-based models (ActuationNet, PotentialNet) demonstrate significantly better adherence to Maxwell's equations compared to the unstructured baselines (DirectNet, DirectGBT). Notably, the PotentialNet exhibits zero curl error, confirming that it satisfies the curl-free condition by construction. The MPEM baseline is excluded from this plot as it inherently satisfies both conditions by construction.}
  \label{fig:CurlDiv}
  \vspace{-0mm}
\end{figure}

\Cref{fig:CurlDiv} shows that, for both systems, the structured models (PotentialNet and ActuationNet) exhibit substantially smaller Maxwell residuals than the direct current-to-field baselines, therefore achieving better physical consistency. Additionally, as intended from its scalar-potential formulation, the PotentialNet successfully produces a curl-free magnetic field model.

\subsection{Evaluation time} 
\begin{figure}[]
  \centering
  \includegraphics{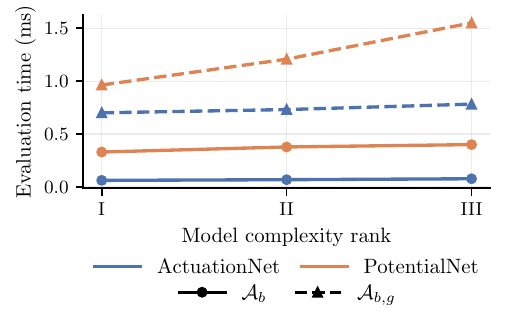}
  \vspace{-7mm}
  \caption{Computational efficiency: Mean evaluation time versus model complexity rank for the structured learning-based models. Solid lines denote the computation of the field actuation matrix ($\actuation_{\field}$), while dashed lines correspond to the joint evaluation of both field and gradient actuation matrices ($\actuation_{\fieldgrad}$). All evaluation times remain below \unit[1.5]{ms}, indicating suitability for high-frequency real-time control. In accordance with real-time control usage, the neural models run on the CPU (Apple M3 (8-core CPU), 16 GB unified memory). The MPEM is not included since, owing to its lightweight closed-form analytical formulation, its evaluation time is negligible on the scale of this comparison.}
  \label{fig:EvalTime}
  \vspace{-2mm}
\end{figure}

Magnetic real-time control hinges on the rapid evaluation of the actuation matrix which needs to be recomputed at each control cycle. We assess the computational efficiency of our control-ready models by averaging runtimes over $10^4$ evaluations (\Cref{fig:EvalTime}). All computed times remain sufficiently small ($<$\unit[1.5]{ms}), confirming that our proposed models are suitable for high-frequency real-time magnetic control.

\subsection{Closed-loop control application}

% To verify the suitability of the proposed models for closed-loop control, we perform a circular trajectory-tracking task with a Lyapunov unstable magnetically actuated rod with two angular degrees of freedom $\alpha$ and $\beta$. Control is implemented using the LQRI methodology described in \cite{ExpandingtheWorkspaceofElectromagneticNavigationSystemsUsingDynamicFeedbackforSingleandMultiagentControl}, running at \unit[100]{Hz} with the considered field models. As shownhttps://www.overleaf.com/project in \Cref{fig:ClosedLoop}, like the MPEM, both proposed models enable successful tracking, demonstrating suitability for real-time inversion.

% \begin{figure}[h]
%   \centering
%   \vspace{-2mm}
%   \includegraphics{figures_revision/trajectory.pdf}
%   \vspace{-5mm}
%   \caption{Closed-loop circular trajectory tracking of the Lyapunov-unstable rod in the ($\alpha$,$\beta$) plane using the MPEM, ActuationNet, and PotentialNet field models. The circular reference trajectory is denoted by a dashed gray line. The results show successful tracking for all three models.}
%   \label{fig:ClosedLoop}
%   \vspace{-5mm}
% \end{figure}

To assess the suitability of the proposed models for closed-loop control, we consider a circular trajectory-tracking task involving a magnetically actuated rod with two angular degrees of freedom, $\alpha$ and $\beta$, following the methodology of \cite{ExpandingtheWorkspaceofElectromagneticNavigationSystemsUsingDynamicFeedbackforSingleandMultiagentControl}. Since the system exhibits inherent Lyapunov instability, achieving stabilization requires a sufficiently high sampling frequency, thereby confirming our models' capacity for real-time inversion. The rod carries a rigidly attached stack of permanent magnets, modeled as a point dipole with magnetic moment $\vect{m}\in\dom{R}{3}$ located at its geometric center, position $\pos{m}\in\dom{R}{3}$. Consequently, both the dipole position $\pos{m}$ and the orientation of the moment $\vect{m}$ are functions of $\alpha$ and $\beta$.

Magnetic torques and forces exerted on the dipole  (i.e. the magnet stack), are given by $\vect{t}=\vect{m}\times\field$ and $\vect{f}=(\vect{m}\cdot\nabla)\field$, respectively, and can be expressed as the linear mapping:
\begin{equation}
    \begin{bmatrix}
    \vect{t}\\
    \vect{f}
    \end{bmatrix}
    =
    \begin{bmatrix}
    \mat{M}_{\field}(\alpha, \beta) & 0 \\
    0 & \mat{M}_{\fieldgrad}(\alpha, \beta)
    \end{bmatrix}
    \begin{bmatrix}
        \field\\
        \fieldgrad
    \end{bmatrix}
    =
    \mat{M}(\alpha, \beta)
    \begin{bmatrix}
        \field\\
        \fieldgrad
    \end{bmatrix},
\end{equation}
where, defining the magnetic dipole moment as $\vect{m} =\left[m_x, m_y, m_z\right]\transposed$,
\begin{equation}
\begin{aligned}
    \mat{M}_{\field}(\alpha, \beta) &= \operatorname{skew}\left(\vect{m}\right)
    =
    \begin{bmatrix}
     0 & -m_z & m_y \\
     m_z & 0 & -m_x \\
     -m_y & m_x & 0
    \end{bmatrix}, \\[0.5em]
    \mat{M}_{\fieldgrad}(\alpha, \beta) &=
    \begin{bmatrix}
        m_x & m_y & m_z & 0 & 0 \\
        0 & m_x & 0 & m_y & m_z \\
        -m_z & 0 & m_x & -m_z & m_y
    \end{bmatrix}.
\end{aligned}
\end{equation}

The resulting torque about the rod's pivot point is $\vect{t}_c=\vect{t} + \tilde{\mat{J}}(\alpha, \beta)\vect{f}$, we refer the reader to \cite{ExpandingtheWorkspaceofElectromagneticNavigationSystemsUsingDynamicFeedbackforSingleandMultiagentControl} for the definition of the mechanical Jacobian $\tilde{\mat{J}}(\alpha, \beta)\in\dom{R}{3\times3}$. By introducing the matrix $\Lambda \coloneqq\<\mat{I} & \tilde{\mat{J}}\>\mat{M}$, where $\mat{I}\in\dom{R}{3\times3}$ denotes the identity matrix, this torque can be expressed as an affine function of the input currents:
\begin{equation}
    \vect{t}_c = \Lambda(\alpha, \beta) \left(
    \begin{bmatrix}
        \mat{A}_{\field}(\pos{m})\\
        \mat{A}_{\fieldgrad}(\pos{m})
    \end{bmatrix}
    \currentsvect
    +
    \begin{bmatrix}
        \field_0(\pos{m})\\
        \fieldgrad_0(\pos{m})
    \end{bmatrix}
    \right).
\end{equation}
Given a desired torque $\vect{t}_c\desired$ computed by the feedback controller, the required setpoint currents $\currentsvect\desired$ are then calculated as:
\begin{equation}
    \currentsvect\desired = 
    \left(
    \Lambda
    \begin{bmatrix}
        \mat{A}_{\field}\\
        \mat{A}_{\fieldgrad}
    \end{bmatrix}
    \right)\pinv
    \left(
    \vect{t}_c\desired - 
    \Lambda
    \begin{bmatrix}
        \field_0\\
        \fieldgrad_0
    \end{bmatrix}
    \right),
\end{equation}
where the dependence on $\alpha$, $\beta$, and $\pos{m}$ has been omitted for compactness and $\dagger$ denotes the Moore-Penrose pseudoinverse.

\begin{figure}[]
  \centering
  \vspace{-2mm}
  \includegraphics{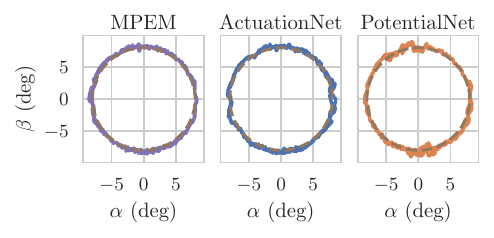}
  \vspace{-5mm}
  \caption{Closed-loop circular trajectory tracking of the Lyapunov-unstable rod in the ($\alpha$,$\beta$) plane using the MPEM, ActuationNet, and PotentialNet field models. The circular reference trajectory is denoted by a dashed gray line. The results show successful tracking for all three models.}
  \label{fig:ClosedLoop}
  \vspace{-5mm}
\end{figure}

Control is implemented using the LQRI methodology described in \cite{ExpandingtheWorkspaceofElectromagneticNavigationSystemsUsingDynamicFeedbackforSingleandMultiagentControl}, running at \unit[100]{Hz} with the considered field models. As shown in \Cref{fig:ClosedLoop}, like the MPEM, both proposed models enable successful tracking, demonstrating suitability for real-time inversion.

\subsection{Workspace ill-conditioning}

\begin{figure*}[!t]
  \centering
  \includegraphics[width=\textwidth]{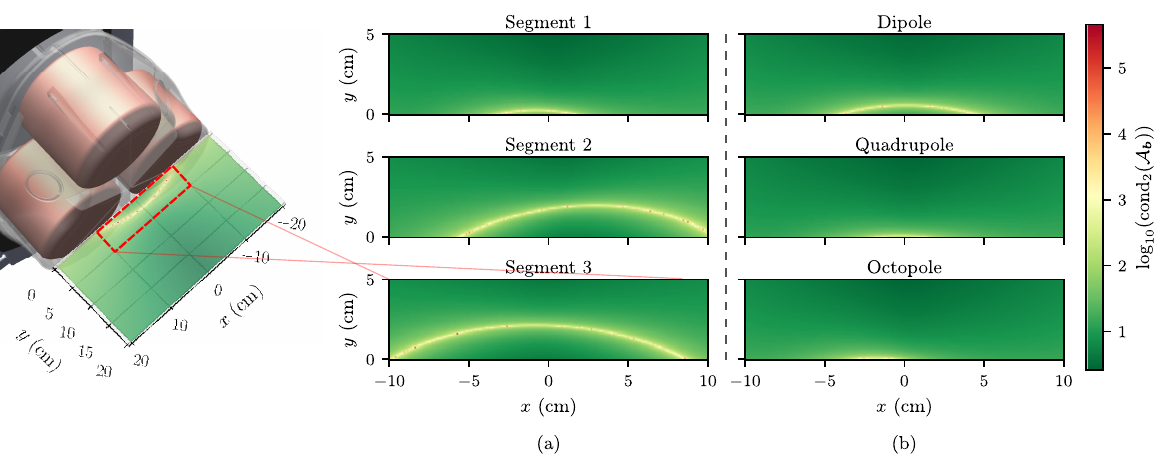}
  \vspace{-9mm}
  \caption{Analysis of workspace ill-conditioning for the MPEM. Heatmaps displaying the logarithmic condition number of the field actuation matrix across a horizontal workspace slice. (a) Effect of training data distribution: Dipole-order MPEMs trained on radial data segments of increasing distance (Segment 1: nearest, Segment 3: farthest). Training on near-field samples (Segment 1) significantly mitigates the ill-conditioned regions (yellow arcs) compared to training on far-field samples. (b) Effect of approximation order: Conditioning maps for MPEMs of increasing expansion order (Dipole, Quadrupole, Octopole), illustrating how increasing the expansion order mitigates ill-conditioning. }
  \label{fig:MpemWorkspace}
  \vspace{-5mm}
\end{figure*}

As reported in \cite{ModelbasedCalibrationForMagneticManipulation, ExpandingtheWorkspaceofElectromagneticNavigationSystemsUsingDynamicFeedbackforSingleandMultiagentControl} and illustrated in \Cref{fig:MpemWorkspace}, the actuation map derived from the MPEM can suffer from significant ill-conditioning. We note that this phenomenon is not exclusive to the MPEM: similar ill-conditioned regions emerge with the PotentialNet, although they are not observed with the ActuationNet. Our analysis focuses on the MPEM, as it serves as the current \textit{de facto} standard model and its ill-conditioning has been explicitly documented in the aforementioned literature.
% \cite{ModelbasedCalibrationForMagneticManipulation, ExpandingtheWorkspaceofElectromagneticNavigationSystemsUsingDynamicFeedbackforSingleandMultiagentControl}.

We observe an empirical trend that calibration samples placed closer to the coils yield markedly better-conditioned actuation maps, whereas emphasizing far-field samples tends to exacerbate ill-conditioned regions. To visualize this behavior, we partition the measured positions of the Navion dataset (\Cref{fig:CombinedMagneticDatasets}(b)) into three radial segments according to their distance, in the horizontal plane, from the Navion electromagnets; the boundary radii are chosen such that all segments contain the same number of positions. We then train a dipole-order MPEM separately on Segment~1 (nearest), Segment~2 (intermediate), and Segment~3 (farthest), and visualize in \Cref{fig:MpemWorkspace}(a) the conditioning of the resulting field actuation map over a horizontal slice of the workspace. Consistent with the observed trend, near-source sampling substantially mitigates the ill-conditioned regions, whereas training predominantly on farther samples exacerbates them. Intuitively, because higher-order terms of the magnetic field decay rapidly with distance from the source, far-field samples are inherently less informative.

Accordingly, we also evaluate the effect of increasing the model's representational capacity by allowing it to capture the contributions of higher-order multipole terms, now using the full Navion dataset. As shown in \Cref{fig:MpemWorkspace}(b), moving from dipole to higher-order MPEM models further reduces ill-conditioning, demonstrating that limited model expressiveness also contributes to spurious conditioning artifacts.

%% file: revision/5-Conclusion.tex
\section{Conclusion}
\label{sec:Conclusion}

In this work, we presented a comprehensive framework for learning-based magnetic field modeling in electromagnetic navigation systems (eMNS), introducing the ActuationNet and PotentialNet architectures. By rigorously benchmarking these data-driven approaches against the physics-based Multipole Expansion Model (MPEM) and unstructured regressors, we demonstrated that learning-based models can effectively replace analytical formulations without sacrificing the algebraic structure required for real-time control.

Our empirical analysis on the OctoMag and Navion systems reveals that the proposed neural models achieve predictive fidelity equivalent to the MPEM, while maintaining comparable data efficiency. Crucially, we identified that the performance of all evaluated models converges to a common RMSE floor for each system. Our sensitivity analysis suggests this limit is governed not by model capacity, but by irreducible geometric uncertainties in the ground-truth acquisition setup. Using the Navion, we also demonstrated that appropriate sampling and sufficient model order effectively eliminate the workspace ill-conditioning frequently reported in MPEM-based calibration.

From a control perspective, this study emphasizes the value of preserving current-linearity. Unlike general non-linear approximations that require iterative optimization for inversion, our affine formulations support closed-form, minimum-norm inversion via the pseudoinverse. This yields evaluation times in the range of approximately \unit[1]{ms}, ensuring that computational latency remains sufficiently low for high-bandwidth feedback loops in real-time magnetic control. Moreover, we experimentally evaluated that the proposed models can indeed support real-time feedback control.

In addition, while analytical models like the MPEM rely on strict assumptions of source symmetry and magnetic isolation, our data-driven models are agnostic to both. This flexibility makes them particularly suitable for applications where idealized analytical assumptions fail---for instance in remote maintenance in nuclear fusion facilities, where complex coil geometries render classical expansions insufficient. 

Finally, the open-source release of our complete codebase and high-density datasets for both clinical- and research-scale systems provides a valuable foundation for future research.

\section*{Acknowledgments}
The authors would like to thank Jonathan Fiene, Felix Grueniger, Arjen Houweling, Rob Faulkner, and Julian Martus for the technical support. Further thanks go to the Max Planck ETH Center for Learning Systems for the financial support. Michael Muehlebach thanks the German Research Foundation for the financial support.